\documentclass[11pt]{article}
\usepackage{graphicx,wrapfig,color}
\usepackage{epsfig}
\usepackage{graphicx}
\usepackage{multicol}
\usepackage{amsfonts}
\usepackage{amsmath,amssymb}
\usepackage{enumerate}
\numberwithin{equation}{section}
\usepackage{cite}

\newcommand{\ben}{\begin{enumerate}}
\newcommand{\een}{\end{enumerate}}
\newcommand{\spa}{\phantom{s}}
\newcommand{\bea}{\begin{eqnarray}}
\newcommand{\eea}{\end{eqnarray}}
\newcommand{\be}{\begin{equation}}
\def\bel#1{\begin{equation} \label{#1}}
\newcommand{\ee}{\end{equation}}
\newcommand{\bi}{\begin{itemize}}
\newcommand{\ei}{\end{itemize}}
\newcommand{\ba}{\begin{align}}
\newcommand{\ea}{\end{align}}

\def\bel#1{\begin{equation} \label{#1}}

\def\vp{\varphi}

\def\mc{\mathcal}
\def\be{\begin{equation}}
\def\ee{\end{equation}}
\def\bea{\begin{eqnarray}}
\def\eea{\end{eqnarray}}

\def\ltap{\ \raise.3ex\hbox{$<$\kern-.75em\lower1ex\hbox{$\sim$}}\ }
\def\gtap{\ \raise.3ex\hbox{$>$\kern-.75em\lower1ex\hbox{$\sim$}}\ }
\def\gl{\ \raise.5ex\hbox{$>$}\kern-.8em\lower.5ex\hbox{$<$}\ }
\def\roughly#1{\raise.3ex\hbox{$#1$\kern-.75em\lower1ex\hbox{$\sim$}}}

\def\nn{\nonumber}

\def\spa{\phantom{a}}
\def\pref#1{(\ref{#1})}

\def\cv{{{\cal{V}}}}

\newcommand\vo{\mathcal{V}}
\newcommand{\comments}[1]{}

\begin{document}

\begin{titlepage}
\today 
\vskip 2 cm
\begin{center}
{\Large \bf  
Inflating Kahler Moduli and Primordial Magnetic Fields}
\vskip 1.5cm  
{ 
{\bf{Luis Aparicio$^{\dagger}$}}, {\bf{Anshuman Maharana$^{*}$}} 
}
\vskip 0.9 cm
{\textsl{$^{\dagger}$Abdus Salam ICTP, \\
Strada Costiera 11, Trieste 34014, Italy. }
}
\vskip 0.5 cm

{\textsl{
*Harish Chandra Research Intitute, \\
HBNI, Chattnag Road, Jhunsi,
Allahabad -  211019, India.\\}
}%

\end{center}

\vskip 0.6cm

\begin{abstract}
We study  the production of primordial magnetic fields 
in inflationary models in type IIB string theory where the role of the 
inflaton is played by a Kahler modulus. We consider various possibilities 
 to realise the Standard Model degrees of freedom in this setting and explicitly determine the time dependence  of the inflaton coupling to the Maxwell term in the models.
Using this we determine the strength and scale dependence of the magnetic fields
generated during inflation. The usual ``strong coupling problem" for primordial magnetogesis manifests itself by
cycle sizes approaching the string scale; this appears in a certain class of  fibre inflation models where the standard model is realised by wrapping D7-branes on cycles in the geometric regime.

\phantom{We study  the production of primordial magnetic fields 
in inflationary models in type II string theory where the role of the 
inflaton is played by a Kahler modulus. We consider various possibilities 
 to realise the Standard Model degrees of freedom in this setting and explicitly determine the time dependence  of the inflaton coupling to the Maxwell term in the models.
Using this we determine the strength and scale dependence of the magnetic fields
generated during inflation. The usual ``strong coupling problem" for primordial magnetogesis manifests itself by
cycle sizes approaching the string scale; this appears in a certain class of  fibre inflation models where the standard model is realised by wrapping D-7 branes on cycles in the geometric regime.Using this we determine the strength and scale dependence of the magnetic fields
generated during inflation. The usual ``strong coupling problem" for primordial magnetogesis manifests itself by
cycle sizes approaching the string scale; this appears in a certain class of  fibre inflation models where the standard model is realised by wrapping D-7 branes on cycles in the geometric regime.The usual ``strong coupling problem" for primordial magnetogesis manifests itself by
cycle sizes approaching the string scale; this appears in a certain class of  fibre inflation models where the standard model is realised by wrapping D-7 branes on cycles in the geometric regime.}

\end{abstract}

\let\thefootnote\relax\footnotetext{electronic address: {\tt{$^{\dagger}$laparici@ictp.it, *anshumanmaharana@hri.res.in} }}

\end{titlepage}
\pagestyle{plain}
\setcounter{page}{1}
\newcounter{bean}
\baselineskip18pt
%



\section{Introduction}
\label{Seckmi}

    Cosmological observations have established a scale invariant and adiabatic spectrum for the inhomogeneities.
This has put the inflationary paradigm as the leading candidate for describing the early universe. In the
next decade, more precision data is expected.  It is hoped that this will help us develop an understanding of the inflationary potential. Two key observables which are likely to play an important
role are the tensor to scalar ratio and non-Gaussianities. Another important aspect is the nature of the coupling
of the inflaton to the Standard Model; this can have various implications for observations. In this context, one
interesting phenomenon is the generation of primordial magnetic fields during inflation \cite{re,1,2,3,4,5,6,7,8,G,9, 11, Seery, mukha}. 

   The scaling symmetry of the Maxwell action implies that there is no production of electromagnetic fields in a FRW background. To generate electromagnetic fields, the scaling symmetry has to be broken. There are two popular
mechanisms for this. The first is the Ratra model which exploits the inflaton  coupling to the Maxwell term in the Lagrangian; the time dependence of the inflaton is then responsible for  breaking the scaling symmetry.  Another avenue is to give the photon a time dependent mass term.  Our focus will be on the former.

    In recent years, there has been a revival in interest in primordial magnetic fields produced during inflation. This has been driven by the search for a mechanism to explain the observed  magnetic fields in various scales, all the way upto the inter-galactic medium \cite{b,r,d}. Although,  there still remain many challenges in obtaining models which are phenomenologically viable (see for e.g. \cite{mukha, Green}).
It was also realised that the generation  of primordial magnetic fields during inflation
provides a source for non-Gaussianities \cite{B, Ba, pele, sloth}. This occurs via a feedback mechanism, the electromagnetic fields produced
act as a source term in the equation governing inflaton fluctuations. In fact, the mechanism also implies correlations
between the magnetic fields produced and the scalar metric perturbations. These satisfy consistency conditions
analogous to the inflationary consistency conditions for scalar three point functions and can serve as a stringent
testing ground for inflation \cite{Jain}. Furthermore, the back reaction of the electromagnetic fields produced can impose constraints on the number of e-foldings during inflation; the strength of this back reaction has to be computed to
check the validity of any model.

   Inflationary dynamics is ultraviolet sensitive via the slow roll parameters. This makes string theory a natural
framework to carry out inflationary model building. In string models
the inflaton coupling to the Maxwell term is set by the gauge kinetic function -- this fixes  the functional
form of the coupling. Apart from the inflationary potential the necessary inputs to compute this coupling are
moduli stabilisation and the scenario for realising the Standard Model sector.   In this paper we will focus on the Large Volume Scenario  (LVS) for moduli stabilisation \cite{LVS, LVS2} in IIB flux compactifications \cite{GKP}.  In IIB flux compactifications the complex structure moduli appear in the Gukov-Vafa-Witten superpotential \cite{gukov}, but Kahler moduli  appear in the superpotential only via non-perturbative effects. The smallness of the non-perturbative effects give an approximately flat potential to the Kahler moduli; they are the candidate inflatons in LVS models. There are two scenarios for inflationary models - blow up \cite{Kahler} and fibre inflation \cite{fibre} models.  We will analyse both the cases. The study is apart of the broader goal to develop a systematic understanding of primordial magneto genesis in string models and see if this can serve as a distinguishing feature between various scenarios.

\section{Review} 

\subsection{Primordial Magnetic Fields From Inflation} 
\label{magrev}

   As discussed in the introduction, scale invariance of the Maxwell action implies no  production of electromagnetic fields in  pure De Sitter space. The symmetry is broken by the inflaton-photon coupling 
\bel{infpho}
S = -{ 1 \over 4} \int d^{4}x I^2(\phi) F_{\mu \nu} F^{\mu \nu}
\ee
Gauge field production is best analysed by expressing the coupling $I^{2}$ in terms of conformal time
(which we will  denote by $\zeta$).  Our discussion shall be very brief, we closely follow  \cite{mukha}. Working in the gauge
$A_0 =0 $, we take the Fourier decomposition of the spatial components of the gauge field to be:
\bel{fd}
  A_{i} (x, \zeta) = {1 \over 2} \sum_{\sigma = 1,2}  \int{ d^3{k} \over (2 \pi)^{3/2}}
  A_{k}^{\sigma} (\zeta)\epsilon^{\sigma}_{i}(k) e^{i\vec{k}. \vec{x}},
\ee
where $\epsilon^{\sigma}_{i}(k)$ are  orthogonal polarisation vectors transverse to the momentum. Defining
\bel{v}
  v_{k}^{\sigma} = \sqrt{ \epsilon^{\sigma}_{i} \epsilon^{\sigma}_{i}} I(\zeta) A_{k}^{\sigma}
\ee
the action reduces \pref{infpho} to
\bel{a}
  S = {1 \over 2} \sum_{\sigma} \int d^3k \bigg( \dot{v}_{k}^{\sigma} \dot{v}_{-k}^{\sigma} - \left[ k^2 - {\ddot{I} \over I}\right] 
  v_{k}^{\sigma} v_{-k}^{\sigma} \bigg),
\ee
where dots indicate derivatives with respect to conformal time. From this we see that the mode functions
are required to satisfy
\bel{required}
    \ddot{v}_{k}^{\sigma} + \omega^2 (\zeta)  {v}_{k}^{\sigma} = 0,
\ee
where $\omega^2 (\zeta) = k^2 - {\ddot{I} \big{/} I}$. For the analogue of the Bunch-Davies vacuum, the
initial conditions for a sufficiently early time $\zeta_{i}$ are 
\bel{ini}
  v_{k}^{\sigma} = { 1 \over \sqrt{ \omega(\zeta_i) } } \spa \spa \spa {\rm and} \spa \spa \spa
  v_{k}^{\sigma} = { i \sqrt{ \omega(\zeta_i) } }
\ee
The power spectrum of the magnetic field can be expressed in terms of $v_{k}^{\sigma}$ and is given by
\bel{bpower}
  \delta_{B} (k, \zeta) = \sum_{\sigma}  { |v_{k}^{\sigma}(\zeta)|^2 k^5
  \over 4 \pi^2 a^4 I^2 }
\ee
The energy density in the electromagnetic field is also determined in terms of the mode functions:
\bel{energy}
  \epsilon = {1 \over 8 \pi a^{4} } \int d^{3}k k^2 \left[ |\dot{v}_{k}^{\sigma}(\zeta)|^2 
  - {\dot{I} \over I} \dot{(|{v}_{k}^{\sigma}(\zeta)|^2)}  + \left( {\dot{I}^2 \over I^2}  +k^2 \right) {|{v}_{k}^{\sigma}(\zeta)|^2} \right]
\ee
We will be interested in the late time behaviour (behaviour at the end of inflation) of the power spectrum of the magnetic field and the electromagnetic energy density. To compute these exactly, the equation for the mode functions has to be solve numerically. But, it is possible to obtain approximate expressions by  matching early and
late time solutions of \pref{required}. The solutions to \pref{required} at early times are plane waves. On the other hand, if the $\ddot{I}/I \gg k^2$ (as is typically
true at late times)   the general solution to \pref{required}  is given by:
\bel{sgen}
  c_1 I (\zeta) + c_2 I (\zeta) \int_{\hat{\zeta}}^{\zeta} { d \rho\over I^2(\rho) }
\ee
The constants $c_1$ and $c_2$ can be (approximately) determined by requiring the early time solution (with initial
conditions as given in \pref{ini}) matches on to the late time solution \pref{sgen} at the time of horizon exit.
Having determined the  constants $c_1$ and $c_2$, one can use \pref{bpower} and \pref{energy} to
obtain the power spectrum of magnetic fields and the energy density at the end of inflation.
\subsection{Large Volume Scenario}
\label{Seclvs}

  As mentioned in the introduction, the models of inflation we will discuss will be in the framework of the
Large Volume Scenario (LVS) for moduli stabilisation \cite{LVS,LVS2}. Here  we briefly review LVS (the reader can find 
a more detailed discussion in the references provided). Moduli stabilisation in LVS is achieved in two steps, first the
stabilisation of complex structure moduli by three form fluxes and then the K\"ahler moduli. The effect of three
form fluxes can be captured by the Gukov-Vafa-Witten  \cite{gukov} super potential
\bel{gvw}
W = \int G_3 \wedge \Omega   \phantom{a}  
\ee
where $\Omega$ is the holomorphic three form of the Calabi-Yau which solely depends on the complex structure moduli.
After integrating out the complex structure moduli, the super potential for the Kahler moduli is given  by
\bel{superi}
 W = W_0 + \sum_{i} A_i e^{-a_i T_i}
\ee
where $W_0$ is the expectation value of the Gukov-Vafa-Witten super  potential, the sum is over the Kahler moduli $T_i = \tau_i +  i c_i$ ($\tau_i$ are the four cycle volumes and $c_i$ their axionic partners). $A_i$ and $a_i$ are constants.
The LVS algorithm can be implemented to stabilise the Kahler moduli if the Calabi-Yau has negative Euler number and 
$W_0$ is of order unity. The stabilisation requires the inclusion of the leading $\alpha'$ correction to the Kahler
potential, it takes the form \cite{BBHL}
\bel{kabab}
K = - 2 \ln \left( \cv + {\hat\xi \over 2} \right),
\ee
where $\hat\xi$ is determined in terms of the Euler number of the Calabi-Yau and vev of the dilaton $\hat\xi = { \chi \over 2 (2 \pi)^3 g_s^{3/2}}$. The simplest models of LVS (which were also the ones to be discovered first \cite{LVS,LVS2}) are the
ones in which the volume of the Calabi-Yau takes  the Swiss-cheese form:
\bel{sw}
  \cv = \alpha \bigg( \tau_1^{3/2} - \sum_{i=2}^{n} \lambda_i \tau_i^{3/2} \bigg).
\ee
The overall volume is controlled by $\tau_1$, the geometric moduli $\tau_2 ,..., \tau_n$ are blow-up modes which can be thought of as parameterising the size of holes in the compactification. The scalar potential in the regime
$\cv \gg 1$ and  $\tau_1 \gg \tau_i \spa ( {\rm{for}} \spa  i >1)$ obtained from the the superpotential \pref{superi} and K\"ahler potential \pref{kabab} is: 
\bel{scalarA}
V_{\rm LVS} = \sum_{i=2}^2 { 8 (a_i A_i)^2 \sqrt{\tau_i} \over 3 \cv \lambda_i } e^{-2 a_i \tau_i}
-  \sum_{i=2}^{n} { 4 a_i A_i W_0 \over \cv^2} \tau_i e^{-a_i \tau_i} + {3 \hat\xi  W_0^2 \over 4 \cv^3 }\,.
\ee 
where the phase of the axionic fields $c_i$ have been adjusted so as to minimise the potential. The potential \pref{scalarA}
has a minimum in the `large volume limit' :
\bel{lim}
\cv \to \infty \spa \spa {\rm{with}} \spa \spa a_i \tau_i \approx \ln \cv\,.
\ee
with a negative value of the cosmological constant. To obtain a minimum with almost vanishing cosmological constant an additional term has to be included in the effective action (for example anti-D3 branes in warped throats \cite{kklt},  dilaton-dependent non-perturbative effects \cite{dil} magnetised D7-branes \cite{mag1, mag2}, or the effect of D-terms \cite{rum}).  The uplift term has the form
\bel{upli}
V_{\rm up} = { D  \over \cv^{\gamma} } \spa \spa \spa {\rm{with}} \spa \spa D > 0\,,
\ee
with  $1\leq\gamma\leq 3$, the precise value of $\gamma$ depends on the  mechanism. 
The coefficient $D$ has to be tuned so that the potential
\bel{total}
V_{\rm LVS} = \sum_{i=2}^2 { 8 (a_i A_i)^2 \sqrt{\tau_i} \over 3 \cv \lambda_i } e^{-2 a_i \tau_i}
-  \sum_{i=2}^{n} { 4 a_i A_i W_0 \over \cv^2} \tau_i e^{-a_i \tau_i} + {3 \hat\xi  W_0^2 \over 4 \cv^3 } + { D  \over \cv^{\gamma} }
\ee
has an approximately Minkowski minimum. 

  In the discussion so far, we have focused on CYs whose volume takes the Swiss cheese form. More generally, the 
volume takes the form 
\bel{genvolu}
  \cv = f (\tau_j) - \sum_k  \beta_k \tau^{3/2}_k,
\ee
where the k-sum runs over the point like blow up moduli and  $f(\tau_j)$ is a homogeneous function of degree $3/2$ 
of the other Kahler moduli (in the limit that the volume of the blow-up moduli is small, the overall volume is given by $f(\tau_j)$). Our interest shall be in the cases in which $f(\tau_j)$ has a fibration structure. To be concrete, consider
a Calabi-Yau whose volume takes the form\footnote{Reference \cite{CKM} provides explicit examples  of this. The base of the fibration is a $\mathbb{P}^{1}$ (the two cycle dual to $\tau_1$) and the fibre is a $K3$ (the four
cycle $\tau_2$)} 
\bel{fib}
   \cv = \alpha \big( \sqrt{\tau_1} \tau_2 - \gamma \tau_3^{3/2} \big).
\ee
Considering the effect of the $\alpha'$ and non-perturbative effects in this setting one finds that  
$\tau_3$ and $\cv$ (which is approximately equal to the product $\sqrt{\tau_1} \tau_2$) are stabilised \cite{Cicoliloops, CKM}. This leaves  one direction in the moduli space unfixed, in order to stabilise this direction
 loop effects have to be considered. The relevant loop effects arise from the exchange
of KK modes and winding modes (see  \cite{berg,wind}  for detailed discussions). Parametrising the unfixed direction by $\tau_1$, inclusion of loop effects, gives a contribution to the moduli potential: 
\bel{Loopcor}
   V_{\rm loop} = g_s^2 { W_0^2 \over \cv^2 } \bigg( {A \over \tau_1^2} + { C \tau_1  \over \cv^2} \bigg) - 
   { W_0^2 \over \cv^2}{ B \over \cv \sqrt{\tau_1} },
\ee
where $A, B, C$ are $\mathcal{O}(1)$ constants.  We will focus on the case where these
constants are positive, since it is this case that gives viable inflationary phenomenology \cite{fibre, robust}. In this case,  one finds that the potential has a minimum at
\bel{min}
   \tau_1 \approx \tau_2 \approx \cv^{2/3}, \spa \spa \spa a_3 \tau_3 \approx \log \cv,
\ee
where $\cv$ is the stabilised value of the overall volume. Note that with this scaling all terms
in \pref{Loopcor} scale as $\cv^{-10/3}$. As before, to obtain an approximately Minkowski vacuum
one has to include an uplift contribution.

     Next, we briefly recall the form of the gauge kinetic functions in realisations of the Standard Model in these 
constructions (see for e.g. \cite{mazu}). This will help us to obtain the inflaton-photon couplings. There are essentially two possible realisations of the Standard model. The first is from D7 branes wrapping four cycles of the Calabi-Yau, in this case the gauge kinetic functions are given by
\bel{geometric}
   f_{i} = {1 \over 4 \pi} T +  h(F_i) S
\ee
where $T$, is the modulus associated with the cycle that the D7 branes wrap. S is the axio-dilaton and $h(F_i)$
are loop suppressed factors which depend on the world volume fluxes. The second is from D3 branes at singularities,
in this case the gauge kinetic functions are
\bel{quiver}
  f_i = S + h T
\ee
where $T$ is the shirking cycle and $h$ a loop factor.

\section{Time Dependence of  Inflaton-Photon Couplings for Inflating \\ Kahler Moduli}

\label{td}

      The basic idea behind inflating Kahler moduli is that inflation occurs when a Kahler modulus
is rolling down to its global minimum in the above described moduli stabilisation scenarios. There
are essentially two possibilities:
\begin{itemize}
\item[(i)]{\it Blow-up Inflation}: The modulus responsible for inflation is a blow up modulus \cite{Kahler}.
\item[(ii)]{\it Fibre Inflation}: The modulus responsible for inflation is a fibre modulus\footnote{Recently, an inflationary scenario where a fibre modulus plays the role of the inflation has been proposed by making use of 
$\alpha'$ corrections to the effective action \cite{SDA}. Our analysis with minor modifications also applies to this case.} \cite{fibre}.
\end{itemize}
\noindent {\underline{Blow-up Inflation:}} 

Here, we collect some aspects of blow-up inflation that will be useful for us later and obtain the expression
for the inflaton-photon coupling in terms of conformal time. We  refer the reader to \cite{Kahler} for details of the model and its phenomenology. For these models, the inflationary dynamics involves one of the blow-up moduli. For concreteness\footnote{For a more general form of the Kahler potential the same analysis can be carried through.} we will take the form of the Kahler potential to be as in \pref{sw}.
Without loss of generality, we take $\tau_n$ to be the modulus displaced from its global minimum. For large values of  $e^{a_n \tau_n}$ ($e^{a_n \tau_n} \gg \cv^2$), one can approximate the potential by \pref{total} : 
\bel{vap}
V_{\rm inf} = \sum_{i=2}^{n-1} { 8 (a_i A_i)^2 \sqrt{\tau_i} \over 3 \cv \lambda_i } e^{-2 a_i \tau_i}
-  \sum_{i=2}^{n-1} { 4 a_i A_i W_0 \over \cv^2} \tau_i e^{-a_i \tau_i}
+ {3 \hat\xi  W_0^2 \over 4 \cv^3 } + { D  \over \cv^{\gamma} } - { 4 a_n A_n W_0 \over \cv^2} \tau_n e^{-a_n \tau_n}\,. \nn
\ee
The inflaton ($\tau_n$) now has an exponentially flat potential, while the other directions remain heavy. Integrating out
these directions one finds the inflaton potential to be
 \bel{ip}
  V_{\rm inf} (\tau_n)  = V_0 - \frac{4
\tau_n W_0 a_n A_n}{\mc{V}_{\rm in}^2}\,e^{-a_n \tau_n} \spa \spa \spa {\rm with} \spa \spa V_{0} = \frac{\b W_0^2}{\mc{V}_{\rm in}^3}\,, 
\ee
where $\cv_{\rm in}$ is the value of the volume during the inflationary epoch and $\beta$ an $\mc{O}(1)$ constant. Phenomenological considerations require $\cv_{\rm in} \approx 10^{5}$.  The field $\tau_n$ is not canonically normalised.  The canonically normalised field in the large volume limit is
\bel{can}
{\sigma} = M_{\rm pl} \sqrt{\frac{4 \lambda_n}{3 \mc{V}_{\rm in}}} \,\tau_n^{\frac{3}{4}}. 
\ee
The inflationary potential in Planck units in terms of $\sigma$ is\footnote{Note that this is similar to an exponential  potential: $V(\sigma)= C_0(1 - e^{-b \sigma})$; with a large coefficient $b$ rather than  large field range for the inflaton.} :
\bel{canpot}
V = V_0 - \frac{4 W_0 a_n A_n}{\mc{V}^2_{\rm in}} \left(\frac{3 \mc{V}_{\rm in} }{4 \lambda_n} \right)^{2/3}  \sigma^{4/3}
\exp \left[-a_n \left(\frac{3 \mc{V}_{\rm in}}{4 \lambda_n}\right)^{2/3} \sigma^{4/3}\right].
\ee
 Before proceeding further, we note K\"ahler moduli inflation suffers from an $\eta$-problem arising as a result of potential loop corrections to the K\"ahler potential.  The issue has to be addressed  by tuning  parameters of the compactification.
 
  In order to obtain the  inflaton-photon coupling as a function of conformal time, one first needs to express the inflaton field excursion in terms of conformal time. To do this, we shall use the standard formula which relates the 
number of e-foldings and the inflaton field excursion 
\bel{fieldfold}
N_e(\sigma) = \int_{\sigma_{\rm end}}^{\sigma}   \frac{1}{\sqrt{2 \epsilon(\sigma)}}\, d\sigma\,.
\ee  
where $\sigma$ is the canonically normalised field, $\epsilon(\sigma)$ the slow roll parameter and $\sigma_{\rm end}$
the field value at which inflation ends. As section \ref{Seclvs}, the in IIB constructions the inflaton-photon coupling is naturally expressed in terms of the field $\tau_n$ (and not the canonically normalised field). Writing
\pref{fieldfold} in terms of the field $\tau_n$ we obtain:
\bel{con}
 \log \bigg( {\zeta \over \zeta_{\rm end} } \bigg) = \frac{3 \b W_0 \lambda_n}{16 \mc{V}_{\rm in}^2 a_n A_n} \int_{\tau^{\rm end}}^{\tau_n}\frac{e^{a_n \tau_n}}{\sqrt{\tau_n}(a_n \tau_n-1)}\, d\tau_n\
\ee
Examining the slow-roll parameters associated with the potential, one finds that inflation ends when $a_n \tau_n = \mathcal{O} (2 \ln \cv_{\rm in}))$, thus in the entire integration of \pref{con}, $a_n\tau_n \gg 1$. Using  this to approximate the integrand, one finds:
\be
 \log \bigg( {\zeta \over \zeta_{\rm end} } \bigg) = \frac{3\b W_0 \lambda_n}{8 \vo_{\rm in}^2 a_n^{3/2} A_n }\left[\frac{e^{a_n \tau_n}}{\sqrt{a_n \tau_n}} +{\rm i} \sqrt{\pi}\, {\rm erf}\left({\rm i}\sqrt{a_n \tau_n}\right)\right]^{\tau_n^{\rm end}}_{\tau_n}\,, 
\label{N-error}
\ee
where ${\rm erf}(x)$ is the error-function. Using  the asymptotic expansion of the error-function for $a_n \tau_n\gg 1 $,
one has
\be
\log \bigg( {\zeta \over \zeta_{\rm end} } \bigg) \approx \frac{3\b W_0 \lambda_n}{16 \vo_{\rm in}^2 a_n^{3/2} A_n }\left[\frac{e^{a_n \tau_n}}{(a_n \tau_n)^{3/2}}\right]^{\tau_n}_{\tau_n^{\rm end}}.
\label{N-approx}
\ee
Let us define
\bel{gam}
  \Gamma \equiv \frac{3\b W_0 \lambda_n}{16 \vo_{\rm in}^2 a_n^{3/2} A_n } \spa \spa \spa {\rm{and}}
  \spa \spa \spa  \theta \equiv \frac{\Gamma e^{a_n \tau^{\rm end}_n}}{(a_n \tau^{\rm end}_n)^{3/2}},
\ee
note that for typical values of the microscopic parameters in Kahler moduli inflation  $\Gamma \sim
\cv_{\rm in}^{-2}\sim 10^{-10}$ while $\theta \sim {\mathcal{O}}(1)$. Using which we write
\bel{gat}
 \frac{e^{a_n \tau_n}}{(a_n \tau_n)^{3/2}} = {1 \over \Gamma} \left[ \log \left( {\zeta \over \zeta_{\rm end}} \right) + \theta \right]
\ee 
The above relation can be inverted using the ProductLog function  to express $\tau_n$ in terms of $\zeta$, giving\footnote{For the definition of the ProductLog we follow the conventions of https://reference.wolfram.com/language/ref/ProductLog.html}:
\bel{pl1}
  a_n \tau_n \approx - { 3 \over 2} \textrm{ProductLog}\left[-1, - {2 \over 3\Gamma^{-2/3}} \left( \log \left( {\zeta \over \zeta_{\rm end}} \right) + \theta \right)^{-2/3}
 \right]
\ee
For large values of $\zeta$ this well approximated by
\bel{invert}
  a_n \tau_n \approx \log \left( {1 \over \Gamma} \left[ \log \left( {\zeta \over \zeta_{\rm end}} \right) + \theta \right]  \right)
\ee

\noindent {\underline{Fibre Inflation:}} 

 Next, let us obtain the analogous expressions for fibre inflation \cite{fibre, robust}. As discussed in section \ref{Seclvs}, the non-perturbative effects and $\alpha'$ corrections stabilise the blow-up cycles and the overall volume but leave the remaining
Kahler moduli unfixed. For the fibration model given in equation \pref{fib} this leaves one unfixed modulus; in fibre 
inflation this direction acts as the inflaton. It is conventional to parametrise this direction by $\tau_1$, 
the canonically normalised inflaton field $\varphi$ is determined  by the relation
\bel{cannon}
   \tau_1  = e^{\kappa \vp}; \spa \spa \spa {\rm with} \spa \spa \spa \kappa = {2 \over \sqrt{3}}
\ee 
Tuning the coefficient of the uplift term such that the cosmological constant
is vanishing at the global minimum and defining $\hat{\varphi}$ by $\hat{\vp} = \varphi -  \langle \vp \rangle$; where $\langle \vp \rangle$ is the value
of $\vp$ at the global minimum\footnote{Reference \cite{fibre} gives $\langle \vp \rangle \simeq {1 \over \sqrt{3}} \log ({ 4A \cv g_s^2\over B } )$}, the inflaton potential can be obtained from \pref{Loopcor}
\bel{Fibpot}
    V( \hat{\vp} ) = { V_0 \over \cv^{10/3} } \left( 3 - 4 e^{-k \hat{\vp}}  - e^{-2 k \hat{\vp} } + R (e^{-k \hat{\vp}} -1)   \right) 
\ee
where $V_0 =  W_0^2\left( { B^{4} \over 256 g_s^2 A} \right)^{1/3} $ and $R = 4 g_s^4 { AC \over B} \ll 1$ \cite{fibre,robust}. For viable inflationary phenomenology $\cv_{\rm in} \lesssim 10^{4}$ for $g_s = 0.1$. With
this potential inflation takes place with $\hat{\vp}$ starting off with a positive value and rolling down to its
minimum. Inflation ends when $\hat{\vp} \simeq \mathcal{O}(1)$. In the region of field space where inflation
occurs the potential is well approximated by
\bel{appropot}
   V( \hat{\vp} ) = { V_0 \over \cv_{\rm in}^{10/3} } \left( 3 - 4 e^{-k \hat{\vp}} \right)
\ee
The relationship between the excursion of the inflaton and  conformal time can be obtain from \pref{fieldfold}; one finds:
\bel{fibefold}
   \log \left( {\zeta \over \zeta_{\rm end} }\right) = \left[ {9 \over 4} e^{\kappa \hat{\varphi} } - \sqrt{3} \hat{\vp} \right]_{\hat{\vp}_{\rm end}}^{\hat{\vp}}.
\ee
Again, it is possible to invert the relationship by using the ProductLog function
\bel{pl3}
   \hat{\vp} = - {1 \over \kappa} \rm{ProductLog} \left[-1, - { 9 \over 4} { \kappa  \over \sqrt{3}} e^{ - \kappa \Omega / \sqrt{3} }  \right] - { \Omega \over \sqrt{3} },
\ee
where we have defined
$$
  \Omega \equiv  \log \left( {\zeta \over \zeta_{\rm end} }\right) + {9 \over 4} e^{\kappa \hat{\varphi}_{\rm end} } - \sqrt{3} \hat{\vp}_{\rm end}
$$
For large values of $\zeta$ the dependence, one finds
\bel{taurel}
   \tau_1 = e^{ \kappa \vp} = \alpha  \log \left( {\zeta \over \zeta_{\rm end} }\right) + \beta
\ee
where $\alpha, \beta$ are $\mathcal{O}(1)$ positive constants determined in terms of A,B and C.

   Recall our discussion in the section \ref{Seclvs}; the inflaton photon coupling is given by $I^2(\zeta) = {{\tau}_{\rm sm} \big{/} 4 \pi}$, where ${\tau}_{\rm sm}$ is the volume of the cycle on which the standard model degrees of freedom are localised . For blow-up inflation the coupling is the the strongest if we take this to be $\tau_1$. Similarly, one can consider realising the standard model on the $\tau_1$ cycle in case of fibre inflation\footnote{ In this case, obtaining a realistic value of the gauge couplings requires mild tuning of the microscopic parameter, as noted in \cite{ZB}. We will not address this issue here.}. The other possibility is to have the standard model degrees of freedom localised on a cycle other than the fibre modulus. Since the volume of the compactification is a constant during inflation, the volume of the cycle associated with the standard model degrees of freedom then scales inversely with the volume of $\tau_1$;  one  has $I^{2}(\zeta) \propto \tau_1^{-p}$ (where $p$ is a positive constant). The value of p depends on the form of the Kahler potential (see \cite{SDA} for various possibilities of the form of the Kahler potential that can lead to fibre inflation). In our analysis in the next section we shall consider both the possibilities. In summary the equations \pref{pl1} and \pref{pl3} can be used to obtain the form of the time dependence of the  coefficient of the Maxwell term  for blow-up and fibre inflation models.

\subsection{Primordial Magnetic Fields from Inflating Kahler Moduli}

   Having obtained the inflaton photon couplings, we are now ready to study magnetic field production during the
inflationary epoch. As described in section \ref{magrev}, to do this we need to compute the late time 
behaviour of the mode functions $v^{\sigma}_{k}$ with  initial conditions as given in  \pref{ini}. Let 
us first compute these using the matching method outlined in section \ref{magrev}  (we will serve as an useful check the numerical we will perform later in the section in the section). In the regime that the term involving $I(\phi)$ can be neglected the solution satisfying the required initial conditions is
\bel{early}
    v_{k}(\zeta) \approx e^{ik(\zeta - \zeta_i)}.
\ee
Let $\zeta =  \zeta_k$ be the time of horizon exit of the mode with wavenumber $k$ $(\zeta_k \approx - {1/k})$\footnote{To be more accurate one can use $\ddot{I(\eta)}/ I(\eta) \simeq k^2$ as the condition for horizon exit}. The constants $c_1$ and $c_2$ are determined by equating \pref{early} and its first derivative with the late time solution \pref{sgen} and its first derivative. This gives
\begin{equation}
      \begin{bmatrix}
        c_1 \\
        c_2
    \end{bmatrix}
    = e^{ik(\zeta - \zeta_k)}\begin{bmatrix}
        \big{(} I'(\zeta_k) \int_{\hat{\zeta}}^{\zeta_k} { d \rho\over I^2(\rho) } + {1 \over I(\zeta_k)} \big{)}
&  \spa  \spa \tiny{-}  I(\zeta_k) \int_{\hat{\zeta}}^{\zeta_k} { d \rho\over I^2(\rho) } 
 \\
       -I'(\zeta_k) & I(\zeta_k)
     \end{bmatrix}
        \begin{bmatrix}
        {1 \over \sqrt{k}} \\
        i \sqrt{k}
    \end{bmatrix} 
\label{sea1}
\end{equation}
Using the above for $c_1$ and $c_2$ and making use of the fact that $c_2$ is independent of $\hat{\zeta}$ we can
write the late time solution as 
\bel{solumukh}
  {e^{ik(\zeta_k - \zeta_i)} \over \sqrt{k}}{ I(\zeta) \over I(\zeta_k)} + e^{ik(\zeta_k - \zeta_i)} \bigg( -{I'(\zeta_k) \over \sqrt{k} } + i \sqrt{k} I(\zeta_k) \bigg) I(\zeta) \int_{{\zeta_k}}^{\zeta} { d \rho\over I^2(\rho) }.
 \ee
 This expresses the magnitude of the late time solution entirely in terms of $k$ and the function $I(\zeta)$. Note that
 this is similar to the form of the solution derived in \cite{9}, but in the above form the integration limits are
 $\zeta_k$ and $\zeta$ (the dependence of the integration limits on the value of the conformal time at the end of inflation has been eliminated). The above form exhibits that the behaviour of at late times is determined properties of the
 integral
\bel{speci}
  \int_{{\zeta_k}}^{\zeta} { d \rho\over I^2(\rho) } \spa,
\ee
in particular on whether the integral is dominated by the upper or lower limit of integration. The functions
$I(\phi)$ for blow-up and fibre inflation were obtained in section \ref{td}.

  Let us now compute the power spectrum of the magnetic field $\delta_B$ as defined in \pref{bpower}. We numerically integrate
\pref{required} over a range of $k$ values (for a particular choice of the values of the microscopic parameters) and then use this to compute $\delta_B$ at the end of inflation. We
exhibit our results for blow-up inflation in Figure 1.
\begin{figure}[h]
  \centering
   \includegraphics[width=0.5\textwidth]{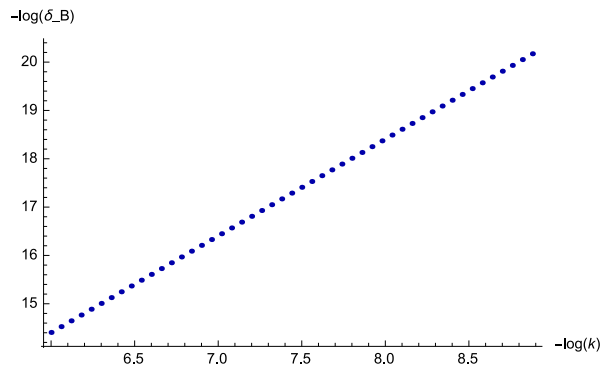} 
 \caption{Power spectrum for a Fibre inflation model with $I^2(\zeta) \propto \tau_1$; with  $\tau^{\rm end}_{\rm sm} = { 4\pi \over g^2} \simeq 25$ }
  \label{fibp power}
\end{figure}
 For this model best fit analysis gives the slope of the power spectrum to  be $s \simeq 2.002$ (with a peak towards higher values of k). For fibre inflation, we consider the  two possibilities discussed at the end of section \ref{td}. Figure 2 exhibits a model with the $I^{2} (\zeta) \propto \tau_1$ (with $\tau^{\rm end}_{\rm sm} = { 4\pi \over g^2} \simeq 25$ at the end of inflation (where the subscript `sm' indicates that the cycle is the one associated with the standard model degrees of freedom)) as required for a realistic value of the gauge coupling). For this, we find $s \simeq 2.02$. 
\begin{figure}[h]
  \centering
   \includegraphics[width=0.5\textwidth]{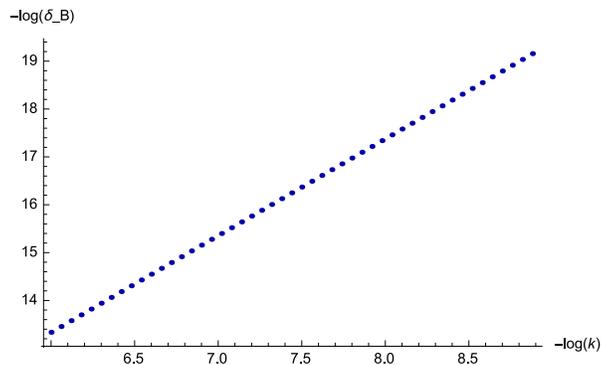} 
 \caption{Power spectrum for a Fibre inflation model with $I^2(\zeta) \propto \tau_1$; with  $\tau^{\rm end}_{\rm sm} = { 4\pi \over g^2} \simeq 25$ }
  \label{fibp power}
\end{figure}
Figure 3 exhibits a model\footnote{This corresponds to $p=1$, as per the notation introduced in section \ref{td}. We will focus on this case, the analysis can be easily generalised to other values of $p$.}  with $I^{2} (\zeta) \propto \tau_1^{-1}$ (with $\tau^{\rm end}_{\rm sm} = { 4\pi \over g^2} \simeq 250$ at the end of inflation). For this, we find $s \simeq 1.97$. 
\begin{figure}[h]
  \centering
   \includegraphics[width=0.5\textwidth]{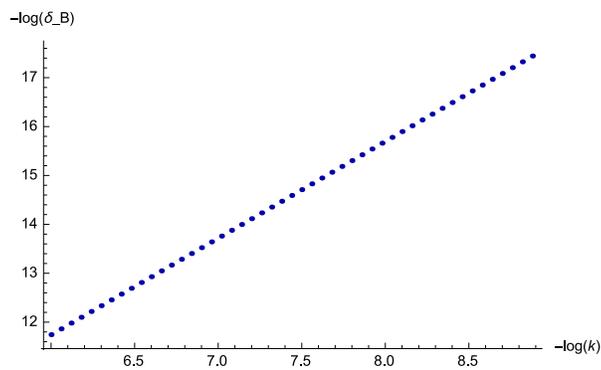} 
 \caption{Power spectrum for a Fibre inflation model with $I^2(\zeta) \propto \tau^{-1}_1$; with  $\tau^{\rm end}_{\rm sm} = { 4\pi \over g^2} \simeq 250$ }
  \label{fibp power}
\end{figure}
Note that in this case the gauge coupling at the end of inflation is one order of magnitude below what is required for realistic model building. For models with $I^{2} (\zeta) \propto \tau_1^{-1}$ and realistic values of the gauge coupling at the end of inflation, numerical evolution (backwards in time) shows that the cycle size falls approaches the string scale during the inflationary epoch before 60 e-foldings are achieved. Hence the effective field theory cannot be trusted. Figure 4 gives the plot of the evolution of the cycle size in this case. This is the manifestation of the ``strong coupling problem"  for primordial magnetogenesis  in this setting. The strong coupling problem is one of the major challenges for obtaining inflationary models of primordial magnetogenesis which can explain the observed magnetic fields in the megaparsec scale (see for e.g. \cite{mukha,Green}). In this setting,  effective field theory breaks down, invalidating the analysis from the point of field theory. Although the effective field theory and breaks downs, there is a new effective field theory which can be used to describe the regime in which the cycle size is well below the string scale (see \cite{ap} for a summary of the form of the Kahler potential and the super potential in the singular regime). 
It will be interesting to explore if this can be used to shed light on the strong coupling problem; we leave this direction for future work (as this requires knowledge of how to describes the system during its transition from the
geometric to the singular regime which at present is not available).
\begin{figure}[h]
  \centering
   \includegraphics[width=0.46\textwidth]{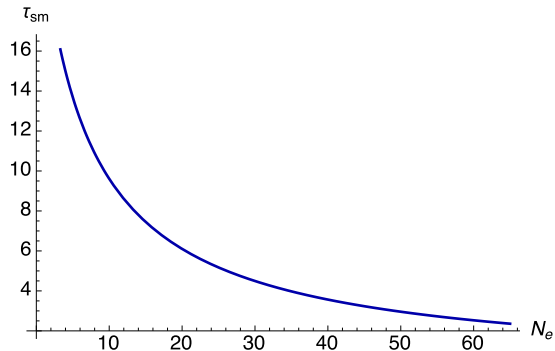} 
 \caption{Evolution of the cycle size backwards in time for a fibre inflation model with $I^{2} (\zeta) \propto \tau_1^{-1}$ and $ \tau^{\rm end}_{\rm sm} = { 4\pi \over g^2} \simeq 25$ at the end of inflation. The cycle size approaches the string scale for $N_e \approx 55$ (the number of e-foldings of the universe); a manifestation of the strong coupling problem.  }
  \label{fibp power}
\end{figure}

 Next, let us obtain the strength of the magnetic field in the present epoch. The magnetic field falls of as $a^{-2}$ with the expansion of the universe.  For fibre inflation the reheating epoch is matter dominated with quanta of the 
fibre modulus, which is followed by a thermal history. The post-inflationary history for blow-up inflation was discussed in detailed in \cite{BLH}. The expansion of the universe after inflation  $(a_0/a_f)$ in both case can be obtained using the discussion in
\cite{D, BLH} (to determine the number of e-foldings that the universe undergoes in the epochs in which the energy density is dominated by moduli quanta) and demanding conservation of entropy after the last epoch of reheating. Using this, for fibre inflation
we find $(a_0/a_f)_{{\rm fib}} \simeq 10^{31}$, on the other hand for blow-up inflation  $(a_0/a_f)_{{\rm bl}} \simeq 10^{30}$. For a model of blow-up inflation with microscopic parameters as in Figure 1, using this and the result of the numerical evaluation described earlier in the section we find the magnitude of the magnetic field at the megaparsec scale   $\delta_B \simeq 10^{-57} G$ for the model.  For the fibre inflation model with microscopic parameters as in Figure 2 the magnetic field in the megaparsec scale is $\delta_B \simeq 10^{-58} G$; while for the model of Figure 3, $\delta_B \simeq 10^{-57} G$. We note that the strength of the magnetic field at the present epoch is well below the observational bounds and  is of the same magnitude as for the racetrack models discussed in \cite{AA}. A more accurate determination
 of the strength of the magnetic fields requires understanding their evolution during reheating (see \cite{rh} for a recent discussion), which is beyond the scope of the present work.

  Finally, let us estimate the back reaction caused from the production of electromagnetic fields in these models.
The contributions of modes of wavelength in the sub horizon scale would be renormalised thus the 
integral in \pref{energy} runs over from the mode to exit the horizon first till the modes exiting the horizon at
the end of inflation. The integral is dominated by the high $k$ modes; this gives the energy density in both cases  to be
$$
   \epsilon \simeq H_I^4
$$
Thus the models are safe from the point of view of back reaction from the production of electromagnetic fields.

\comments{
 Rough estimate of the
integral \pref{speci} then gives that for both the models
\bel{esti}
  |v^{\sigma}_k (\zeta)| \propto {1 \over \sqrt{k}}{ I(\zeta) \over I(\zeta_k)},
\ee
where we have used the fact that $\zeta_k \simeq -{1 / k}$. To check the validity of the matching solutions we compare with numerical results. Figure xxx show plots  of these solutions for a typical choice of parameters
for fibre inflation. Note that the solutions are in agreement at late times.
}

\comments{

\section{To do}

\begin{itemize}

\item Value of $I_{\rm end}$

\item Inverse dependence

\item $\tau_1 = { 4\pi \over g^2} = 25$

\item giovanni

\end{itemize}
}
\section{Conclusions}
  
   We have analysed primordial magnetic field production for inflating Kahler moduli in the large volume scenario
for moduli stabilisation by considering various possibilities for realising the Standard Model sector. The study is part of the broader goal of trying to develop a systematic understanding of
the generation of primordial magnetic fields in string models.  We determined the scale dependence and strength of the power spectrum - $\delta_B$. The strength of the magnetic fields in the present epoch is found to be well below the observational bounds. For fibre inflation models in which the standard model cycle shrinks during inflation, there is a strong coupling
problem. We found that geometrically this corresponds to a cycle size becoming less than the string scale -- it will be interesting to explore if the effective field theory in the singular regime can shed any light on the issue. Since fibre inflation models predict a higher value of the tensor to scalar ratio in comparison to most string models, it will be interesting to explore whether the ``strong  coupling problem"  arises in other string models which predict a high value of $r$ to see if this is a generic feature. Finally, we have estimated the energy density produced and found that  this does not lead to any back reaction problems for the models.

\section*{Acknowledgements}

We would like to thank Koushik Dutta for discussions.  AM is partially supported by a Ramanujan Fellowship. AM would like to thank ICTP, Trieste for hospitality.

\end{document}